\documentclass[fleqn,usenatbib]{mnras}

\usepackage{newtxtext,newtxmath}
\usepackage[T1]{fontenc}
\usepackage{float}

\DeclareRobustCommand{\VAN}[3]{#2}
\let\VANthebibliography\thebibliography
\def\thebibliography{\DeclareRobustCommand{\VAN}[3]{##3}\VANthebibliography}

\usepackage{graphicx}	% Including figure files
\usepackage{amsmath}	% Advanced maths commands
\title[Galaxy groups -- cosmic web nodes]{The localization of galaxy groups in close proximity to galaxy clusters using cosmic web nodes}
\author[D. Cornwell et al.]{\parbox{\textwidth}{
  Daniel J. Cornwell$^{1}$\thanks{E-mail: daniel.cornwell@nottingham.ac.uk}, 
  Ulrike Kuchner$^{1}$, 
  Meghan E. Gray$^{1}$, 
   Alfonso Arag\'{o}n-Salamanca$^{1}$, 
  Frazer R. Pearce$^{1}$,
  Weiguang Cui$^{2,3,4}$,
  Alexander Knebe$^{2,3,5}$
  } 
\\
\\
$^{1}$School of Physics and Astronomy, University of Nottingham, Nottingham NG7 2RD, UK\\
$^{2}$Departamento de F\'isica Te\'{o}rica, M\'{o}dulo 15, Facultad de Ciencias, Universidad Aut\'{o}noma de Madrid, 28049 Madrid, Spain\\
$^{3}$Centro de Investigaci\'{o}n Avanzada en F\'isica Fundamental (CIAFF), Facultad de Ciencias, Universidad Aut\'{o}noma de Madrid, 28049 Madrid, Spain\\
$^{4}$Institute for Astronomy, University of Edinburgh, Royal Observatory, Edinburgh EH9 3HJ, United Kingdom\\
$^{5}$International Centre for Radio Astronomy Research, University of Western Australia, 35 Stirling Highway, Crawley, Western Australia 6009, Australia
}

\date{Accepted XXX. Received YYY; in original form ZZZ}

\pubyear{2015}

\begin{document}
\label{firstpage}
\pagerange{\pageref{firstpage}--\pageref{lastpage}}
\maketitle

\begin{abstract}
We investigate the efficacy of using the cosmic web nodes identified by the DisPerSE topological filament finder to systematically identify galaxy groups in the infall regions around massive clusters. The large random motions and infall velocities of galaxies in the regions around clusters complicate the detection and characterisation of substructures through normal group-finding algorithms. Yet understanding the co-location of galaxies within filaments and/or groups is a key part of understanding the role of environment on galaxy evolution, particularly in light of next-generation wide-field spectroscopic surveys. Here we use simulated massive clusters from TheThreeHundred collaboration and compare the derived group catalogues, (haloes with $\sigma_{v} > 300  h^{-1}$ km/s) with the critical points from DisPerSE, ran on haloes with more than 100 particles. We find that in 3D, 56\% of DisPerSE nodes are correctly identified as groups (purity) while 68\% of groups are identified as nodes (completeness). The fraction of matches increases with group mass and with distance from the host cluster centre. This rises to a completeness of 100\% for the most massive galaxy groups ($M>10^{14}$ M$_{\odot}$) in 3D, or 63\% when considering the projected 2D galaxy distribution. When a perfect match occurs between a cosmic web node and a galaxy group, the DisPerSE node density ($\delta$) serves as an estimate of the group’s mass, albeit with significant scatter. We conclude that the use of a cosmic filament finder shows promise as a useful and straightforward observational tool for disentangling substructure within the infall regions of massive clusters.

\end{abstract}


%%%%%%%%%%%%%%%%%%%%%%%%%%%%%%%%%%%%%%%%%%%%%%%%%%

%%%%%%%%%%%%%%%%% BODY OF PAPER %%%%%%%%%%%%%%%%%%

\section{Introduction}
The cosmic web is a vast network connecting the matter in the Universe \citep{Bond96}, made up of voids, sheets, filaments and nodes. This structure arises due to the presence of small perturbations that propagate through the early Universe's primordial plasma resulting in over and underdensities, providing the seeds of structure growth \citep{Springel06}. Over cosmic time, these fluctuations are amplified through gravity and build highly asymmetrical structures. Overdense regions firstly collapse to form walls, then collapse through two principal axes to form filaments before finally forming clusters \citep{Arnold81}. Galaxies follow the large-scale distribution of dark matter and can therefore be used as tracers for the cosmic web.

Some galaxies exist in galaxy cluster sized dark matter haloes. Such galaxies are subject to frequent interactions, both with neighbouring subhaloes and satellites, as well as with the intracluster medium. The impact of the environmental density on the properties of a galaxy can be clearly seen in the morphology density relation: at greater environmental densities, there is a higher fraction of early type galaxies \citep{Dressler_80}. Beyond the virial radius, galaxies are fed into the cluster via cosmic filaments and/or groups \citep{Florian, Mart_nez_2015}, where they experience "pre-processing" \citep{Zabludoff_1998}. Whilst the extent and location of pre-processing is still debated, recent studies at low and intermediate redshifts have shown that galaxies experience this effect before they reach first infall \citep{Tawfeek_2022, werner2022satellite}, providing motivation for the study of the influence of filaments and groups on galaxy evolution. Next generation wide-field, multi-object spectroscopic surveys, such as the WEAVE Wide Field Cluster Survey \citep[WWFCS;][]{Jin} and the 4MOST CHileAN Cluster galaxy Evolution Survey (CHANCES; Haines et al. in prep) will directly address the need for this study. By obtaining thousands of galaxy spectra out to several virial radii around low-redshift clusters, these surveys will investigate the impact of the cosmic web around galaxy clusters and the properties of the galaxies that lie within it.

In preparation for the WWFCS, multiple studies have investigated the feasibility of detecting the cosmic web around galaxy clusters, using galaxies as tracers for the underlying gas and dark matter skeleton \citep{Kuchner20,Kuchner21,Kuchner22,Cornwell_2022, Cornwell_23}. 

To understand to what extent the cosmic web influences galaxy evolution in the infall region around galaxy clusters, a method for robustly detecting galaxy groups is needed. Currently, there exists a range of techniques that successfully detect groups in larger-scale observations, namely the commonly used Friends-of-Friends (FoF) percolation algorithm \citep{FoF}, which has been used, for example in the SDSS \citep{Berlind_2006}, GAMA \citep{Robotham_2011} and 2dFGRS \citep{Eke04}. However, in the infall region of galaxy clusters, group-finding is a non-trivial task. For example, inside the cluster's potential well, galaxies have large random motions relative to one another. In addition, infall motions towards the cluster and filaments also dramatically perturb the galaxy distribution \citep{Kuchner21}. Therefore, with respect to the observer, galaxies in the vicinity of clusters and groups may have similar distances, but their large random motions lead to very different redshifts. This manifests as long, artificially extended structures, known as the `Fingers of God' \citep[FoG;][]{Tully_fisher_1978}. The length of the FoG for a massive cluster with velocity dispersion of $1400$ km s$^{-1}$ corresponds to $20\  h^{-1}$ Mpc extending in each direction \citep{Kuchner21}. Therefore, in the vicinity of galaxy clusters, we are limited to more laborious, non-systematic methods of group detection that are used on a cluster-by-cluster basis. For example, previous studies have relied on visually inspecting 3D maps of the galaxies in RA, DEC, and redshift space to detect possible galaxy overdensities \citep{Jaffe13} or the Dressler-Shectman test that compares the local velocity and velocity dispersion for each galaxy against a global value \citep{DS}. These and other methodologies are successful on a single cluster basis but become very time consuming when considering multiple clusters and therefore hundreds of groups are observed. It would prove beneficial to develop a systematic way of reliably detecting galaxy groups in and around clusters.

Cosmic web nodes denote areas in the large scale distribution where filaments intersect. They generally align with peaks in the density field which signal the presence of massive haloes, typically representing clusters or galaxy groups.
With this in mind, \cite{Cohn22} used \texttt{DisPerSE}, a topological structures extractor \citep{Sousbie11, Sousbie_2_11}, to test the matching of the location of galaxy cluster-sized haloes ($M_{200} > 10^{14} h^{-1}\ M_{\odot}$) to cosmic web nodes in the  Millennium simulation \citep{Millennium}. Using a variety of input network parameters and matching techniques, they found that $75\%$ of galaxy clusters are matched to a \texttt{DisPerSE} node, implying that galaxy clusters represent peaks in the cosmic web. Furthermore, \citep{galárragaespinosa2023evolution} fine-tuned their \texttt{DisPerSE} input parameters based off the matching of peaks in the Delaunay density field to massive haloes. Both of these studies were performed on cosmological box scales. However, it is unknown whether the matching of nodes to high-mass haloes extends to group-sized haloes and to scales comparable to that of the WWFCS, (regions encompassing galaxy cluster outskirts, typically out to $5R_{200}$). Furthermore, the complexity of the infall region of galaxy clusters, being the interface between the dense, non-linear cluster core and the larger-scale cosmic web, adds significant complexity to this matching. 

Motivated by upcoming wide-field observations of galaxy clusters, in this paper, we investigate the reliability of using \texttt{DisPerSE} to systematically locate galaxy groups in and around clusters by the simple process of identifying nodes in the filament network. The motivation is to encapsulate the individual components of the cosmic web (the clusters, groups and filaments) together, as one evolving field.

This paper is organized as follows: in Section~\ref{sec:Data}, we describe the simulation data used in this project. In Section~\ref{sec:ground_truth_groups}, we introduce our reference galaxy groups and describe the identification of cosmic web nodes. In Section~\ref{sec:Results} we interpret the outcome of matching cosmic web nodes to galaxy groups. In Section~\ref{sec:group_mass_node_dens} we discuss the feasibility of using cosmic web node densities to interpret the mass of galaxy groups. Finally, in Section~\ref{sec:Conclusions}, we present our conclusions.

\section{Data catalogues}
\label{sec:Data}
The analysis presented here is based on \textsc{TheThreeHundred}\footnote{\url{https://the300-project.org/}} galaxy cluster project \citep{Cui2018, Cui22}. Briefly, \textsc{TheThreeHundred} is a set of 324 zoom-in resimulations of the most massive galaxy clusters, identified at $z=0$ in the parent Multidark (MDPL2) simulation \citep{Klypin2016}. MDPL2 is a periodic cube of comoving length $1\,h^{-1}\,$Gpc containing $3840^3$ dark matter particles, each with mass $1.5 \times 10^{9}\,h^{-1} {M_{\odot}}$. MDPL2 uses \textit{Planck} \citep{Planck15} cosmology ($\Omega_{\text{M}} = 0.307,  \Omega_{\text{B}} = 0.048,  \Omega_{\Lambda} = 0.693,  h = 0.678,  \sigma_8 = 0.823,  n_s = 0.96$). 

\textsc{TheThreeHundred} project locates the 324 most massive haloes ($M_{\text{200}} > 8 \times 10^{14} h^{-1}M_{\odot}$) at $z=0$. The dark matter particles within a $15\,h^{-1}$ Mpc sphere around these objects at $z=0$ are traced back to the initial timestep. The highest resolution dark matter particles are then split into dark matter and gas, following the cosmological baryonic mass fraction using the Planck 2015 cosmology $m_{\text{DM}} = 1.27 \times 10^{9} h^{-1} M_{\odot}$ and $m_{\text{gas}} = 2.36 \times 10^{8} h^{-1} M_{\odot}$. The cluster is then resimulated using multiple hydrodynamic physics codes, with additional low-resolution particles retaining the influence of the tidal field of the cosmological structure outside the resimulated volume. We use the zoom re-simulations using \textsc{Gadget-X} which incorporate full-physics galaxy formation, star formation and feedback from both SNe and AGN.  The work in this paper utilizes the AMIGA Halo Finder  \citep[AHF;][]{Gill2004,Knollmann2009} to determine the halo positions and properties and we take the redshift $z=0$ snapshot which is comparable to the low redshifts of the WWFCS clusters.

In this paper, we use a sub-sample of simulated galaxy clusters from \textsc{TheThreeHundred} that we assembled in \cite{Cornwell_2022}. This particular sample was selected in order to create mock observations that were appropriated matched to the WWFCS clusters \citep{Jin}. There are 10 simulated galaxy clusters mass-matched to each of the 16 identified WWFCS clusters for a total of 160 cluster simulations. They populate a cluster mass range of $ 13.8 < \text{log}_{10} M_{200} < 15.2$, all identified at the $z=0$ snapshot. For each galaxy cluster, we take all of the haloes identified by the AHF that exceed a halo mass of $1.5 \times 10^{11} h^{-1} M_{\odot}$ which corresponds to the accumulative mass of 100 high-resolution dark matter particles. This cut lies well above the stellar mass limit of the WWFCS, \citep{Jin}. We assume that in the real observations, every halo that we use in this analysis will host a galaxy.

Motivated by the WWFCS (see \citealt{Cornwell_2022,Cornwell_23} for details), we carry out the analysis in this paper using both the full 3D cluster simulations and the 2D-projected cluster data: i.e., the \textit{3D simulated clusters} correspond to the full 3D cluster region, using the $x$, $y$ and $z$ positions of the simulated haloes; and 
the \textit{2D projections} correspond to the same clusters projected into 2D, using the $x$ and $y$ positions and omitting the $z$ component. \cite{Kuchner21} showed that it isn't currently feasible to reconstruct filamentary networks in 3D in surveys such as the WWFCS, hence why we project the networks in 2D for this approximation.

\begin{figure*}
    \centering
    \includegraphics[width = \textwidth]{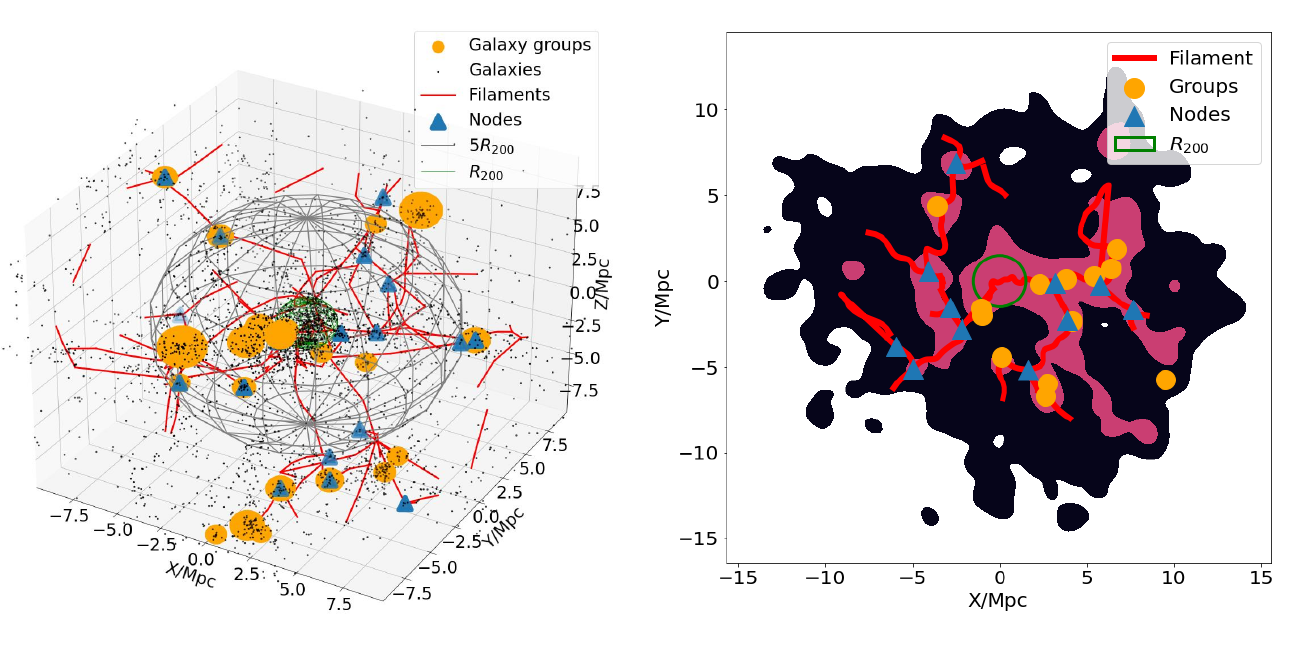}
    \caption{A mass-matched simulated cluster analogue of WWFCS cluster RX0058 ($M_{200} = 4.3 \times 10^{14} \text{M}_{\odot}$). It corresponds to Cluster 237 from \textsc{TheThreeHundred}. Left: the full 3D simulated galaxy cluster. Galaxy haloes are plotted as black dots and the 3D filament network is plotted as red lines. The larger grey sphere corresponds to $d_\text{3D} = 5R_{200}$; the smaller orange spheres represent galaxy groups, and the central green sphere illustrates the cluster core ($d_\text{3D}=R_{200} = 1.3$ Mpc). The nodes of this particular network outside $1.5R_{200}$ are shown as blue triangles. Right: an illustration of the 2D projection of the same galaxy cluster. A kernel density estimate (KDE) is used to represent the halo density distribution with a smoothing scale of $500\,$kpc. The positions of the `true' galaxy groups and the nodes identified by \texttt{DisPerSE} are shown as orange circles and blue triangles, respectively. The filament network is shown by the red lines. The green circle corresponds to $d_\text{2D}=R_{200}$.}
    \label{fig:example_cluster}
\end{figure*}
\section{Identifying the large-scale structure around galaxy clusters}
\label{sec:ground_truth_groups}
In the following section we discuss the identification of galaxy groups and cosmic web networks around our simulated galaxy clusters.

\subsection{Ground truth galaxy groups}
In order to address the main science question in this paper, whether cosmic web nodes coincide with the location of galaxy groups, we firstly need to identify the `true' galaxy groups.

 Motivated by the forthcoming WWFCS observational strategy \citep{Kuchner22}, we identify group centres in the \textsc{TheThreeHundred} simulations using haloes contained in the region $1.5R_{200}<d<5.5R_{200}$, where $d$ is the radial distance to the cluster centre. We stay well clear of the complex cluster core regions, where the peculiar velocity and the converging filament networks makes the identification of groups difficult and unreliable. This also ensures that the selected volume for all clusters is completely contained within the high resolution region and also follows the scales that the WWFCS will probe. The group haloes are then selected as objects with 1D velocity dispersion \mbox{$\sigma_{v} > 300\  h^{-1}$} km/s (this is derived from the subhalo velocities and corresponds to a halo mass of $\sim 10^{13} \text{M}_{\odot}$). In the \textsc{TheThreeHundred} database all galaxies within a sphere of $R_{200}$ of the group halo are labelled as group galaxies. However, in this work, we call `groups' the individual haloes that exceed this mass/velocity dispersion threshold, and are not concerned with group subhaloes (or group members).

In the 160 simulated galaxy clusters that we study there are 1775 galaxy groups in the 3D simulated clusters and 2430 in the 2D projections. The group catalogue used in the 2D projections is the same as the 3D group catalogue but is projected onto x and y positions. The difference in the number of groups in the 3D sample and the 2D sample stems from restricting to $1.5R_{200} < d_{\text{3D}} < 5.5R_{200}$ in 3D (volume) and $1.5R_{200} < d_{\text{2D}} < 5.5R_{200}$ in 2D (surface area). In the 2D sample, there are background and foreground group interlopers.

\subsection{Cosmic web networks}
\label{sec:Cosmic_filaments}
We make use of the widely used structures extractor algorithm \texttt{DisPerSE} \citep{Sousbie11, Sousbie_2_11} to identify filaments in the simulated clusters. \texttt{DisPerSE} identifies persistent topological features in an underlying density field, such as peaks, walls, voids and, in particular, filamentary networks. Firstly, the density is derived from the Delaunay tessellation of a discrete particle distribution. Then, \texttt{DisPerSE} computes the Morse-smale complexes \citep{Morse} and extracts the filamentary networks from the critical points: maxima, saddle points, and minima. Nodes are identified as the maxima. Arcs linking maxima to saddle-points trace the filamentary structures whilst ascending/descending manifolds map the voids and walls. These combine to form a filamentary skeleton that trace the topologically significant regions in the density field. 

\texttt{DisPerSE} includes user-input parameters, namely \textit{persistence} and \textit{smoothing}. Persistence is defined as the ratio of the density value of a pair of topologically significant critical points. A pair of critical points form a persistence pair and is accepted or rejected from the network based on the underlying persistence threshold $\sigma$. Essentially, the persistence dictates the robustness of identified structure. The second input parameter defines the degree of smoothing, which determines the straightness of the paths the filaments trace. Previously, we have discussed the importance of mass weighting  to find nodes and robust filaments using the same simulations \citep{Kuchner20, Cornwell_23}. Here, we apply this mass-weighting for our main analysis, but also consider the non mass-weighted networks for comparison.

In this paper, we aim to test the matching of \texttt{DisPerSE} nodes to galaxy groups close to massive galaxy clusters. We run \texttt{DisPerSE} on the haloes that exceed the mass cut of $M > 1.5 \times 10^{11} h^{-1} M_{\odot}$. We firstly perform this test on our 3D filament networks that are ran on the 3D simulated clusters. Then, we aim to extend this to our 2D projections by running the filament networks on the 2D projected clusters (clusters projected in 2D), described in Section~\ref{sec:Data}. 

\subsubsection{3D filament networks}
\label{sec:3D_filaments}
In order to retrieve consistent, representative filament networks with \texttt{DisPerSE}, we need to decide on the input parameters. We choose to approximately match the number of nodes to the number of groups so that we can make a direct comparison between the two. To evaluate the matches, we define completeness and purity in the following way:
\begin{equation}
    \text{Purity} = \frac{\text{Number of nodes matched to groups}}{\text{Number of nodes}},
\end{equation}
\begin{equation}
    \text{Completeness} = \frac{\text{Number of groups matched to nodes}}{\text{Number of groups}},
\end{equation}
which can be computed for the 3D reference simulations and the 2D projections. In our context, the purity we calculate is used to answer: ``when we find a node, how often is it truly a group?'' For comparison, the completeness can be thought of as: ``of all the groups that exist, how many can we find just by identifying nodes?'' For example, if we use a low persistence, it is possible we will have many more nodes than groups. This will result in a high completeness but a low purity as there will be more nodes that have the potential to match to a group. By approximating the number of groups to the number of nodes, we avoid making the choice of maximizing the purity or the completeness.

For the 3D simulated mass-weighted networks, we use a persistence of $3.3\sigma$. We use a smoothing of 5, following \cite{Cornwell_23}, and note that smoothing does not significantly alter the positions of the critical points. To avoid a bias in our statistics, (for example, matching multiple nodes to a group), we further clean the filament networks by omitting any cosmic web node that is within 0.5 Mpc of another node, which is the case for approximately 15 nodes for the whole cluster sample (i.e., less than 1\%). In this process we keep the node with the highest density field value ($\delta$), which is the density contrast computed in the Delaunay tessellation (see below). Summed over all 160 clusters, there are 1818 \texttt{DisPerSE} nodes at clustercentric distances $1.5R_{200} < d_{\text{3D}} < 5.5R_{200}$, compared to 1775 groups. 

\subsubsection{2D filament networks}
In order to make a direct link to observations (see \citealt{Kuchner21} and \citealt{Cornwell_2022} for details), we produce filament networks using the 2D projections of the simulated galaxy clusters.  We run \texttt{DisPerSE} on the $x$ and $y$ positions of the haloes and apply mass-weighting to construct the 2D projected filament networks. Here, we use a persistence of $2.7\sigma$ and a smoothing of 5. After cleaning the networks in the same manner as described above, we produce 2327 nodes, compared to 2430 groups in the range $1.5R_{200} < d_{\text{clus,2D}} < 5.5R_{200}$. 

We note that when observing real cluster regions we don't know \textit{a priori} the true number of groups, which we have used to set the persistence value. As we have done here, one can use simulations to estimate the expected number of groups for clusters of a given mass, and use that to set the persistence value to derive the observed filament network.

An example simulated galaxy cluster is shown in Figure~\ref{fig:example_cluster}. In the left panel, we see the full 3D cluster with the filament network (derived using mass-weighting) overlaid in blue. Galaxy groups are illustrated as orange spheres where their radius corresponds to $R_{200}$ of the group itself. Cosmic web nodes are shown as blue triangles. We show the corresponding 2D projection of the same cluster in the right panel. Here, a kernel density estimate is used to display the cluster density field, and the filament networks are plotted in red. As before, we show the positions of the galaxy groups and cosmic web nodes as yellow circles and blue triangles.
\subsection{Node and group number densities}

\begin{figure}
    \centering
    \includegraphics[width = 0.48\textwidth]{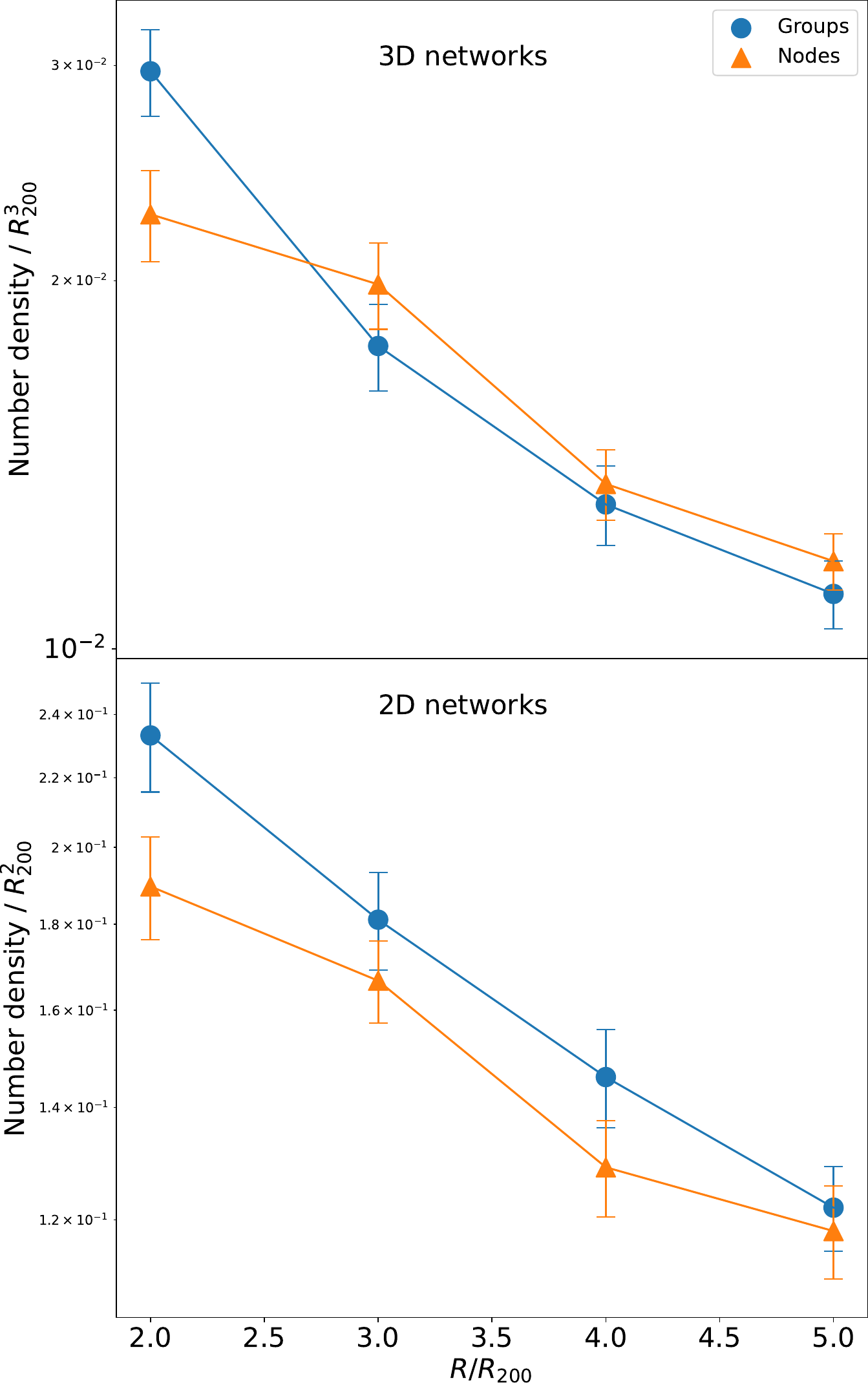}
    \caption{The radial number density of galaxy groups and cosmic web nodes decrease as a function of cluster distance. The top panel shows results from mass-weighted \texttt{DisPerSE} 3D filament networks, the bottom from projected 2D networks. The points show the mean number densities and the error bars are the corresponding standard errors. We find the largest discrepancy between nodes and groups at small clustercentric distances.}
    \label{fig:radial_groups_nodes}
\end{figure}
To test whether \texttt{DisPerSE} nodes match with galaxy groups around clusters, we first compare their number densities as a function of radius,  both in 3D and in projected 2D.
To do this, we calculate the volume and surface number density of groups and nodes in a range of concentric shells for each cluster, where the volume and surface number densities are calculated in units of $(r_{\rm 3D}/R_{200})^{-3}$ and $(r_{\rm 2D}/R_{200})^{-2}$ respectively.

The number density of groups and nodes follows a monotonic decline with clustercentric distance both in 3D and 2D (Figure~\ref{fig:radial_groups_nodes}). 
Close to clusters, the number density of nodes and groups are different, both in 3D and 2D: in the innermost radial bins we identify significantly fewer nodes per unit volume (area) than groups.  We therefore expect that this mismatch may affect the completeness of the matching of nodes to groups near the cluster cores. This discrepancy is especially relevant in light of our initial decision to approximately align the total numbers of nodes and groups. We interpret this as due to the cluster core dominating the local density field and thereby diminishing the likelihood of persistence pairs forming close to the cluster core.
Beyond $2.5 R_{200}$, the number density of nodes and groups begin to converge and agree within each other's standard error. We discuss the implications of these results in more detail in the following section.

\section{Results}

Motivated by surveys such as the WWFCS, we investigate whether cosmic web nodes as detected with \texttt{DisPerSE} match to galaxy groups in the outskirts of galaxy clusters. We carry out this analysis with haloes in cluster simulations in 3D and in projected 2D.
\label{sec:Results}

\subsection{Matching groups to nodes in 3D simulations}
\label{sec:Nodes_3D}

To test the coincidence between nodes and groups, we compute the nearest neighbour from every node to every group in each individual cluster. For there to be a successful match, we require that the node be within a radial distance of $R_{200}$ of the corresponding group centre. Where there are matches between multiple nodes and groups, we take the node with the highest density as calculated by \texttt{DisPerSE}. 

\begin{figure*}
    \centering
    \includegraphics[width = \textwidth]{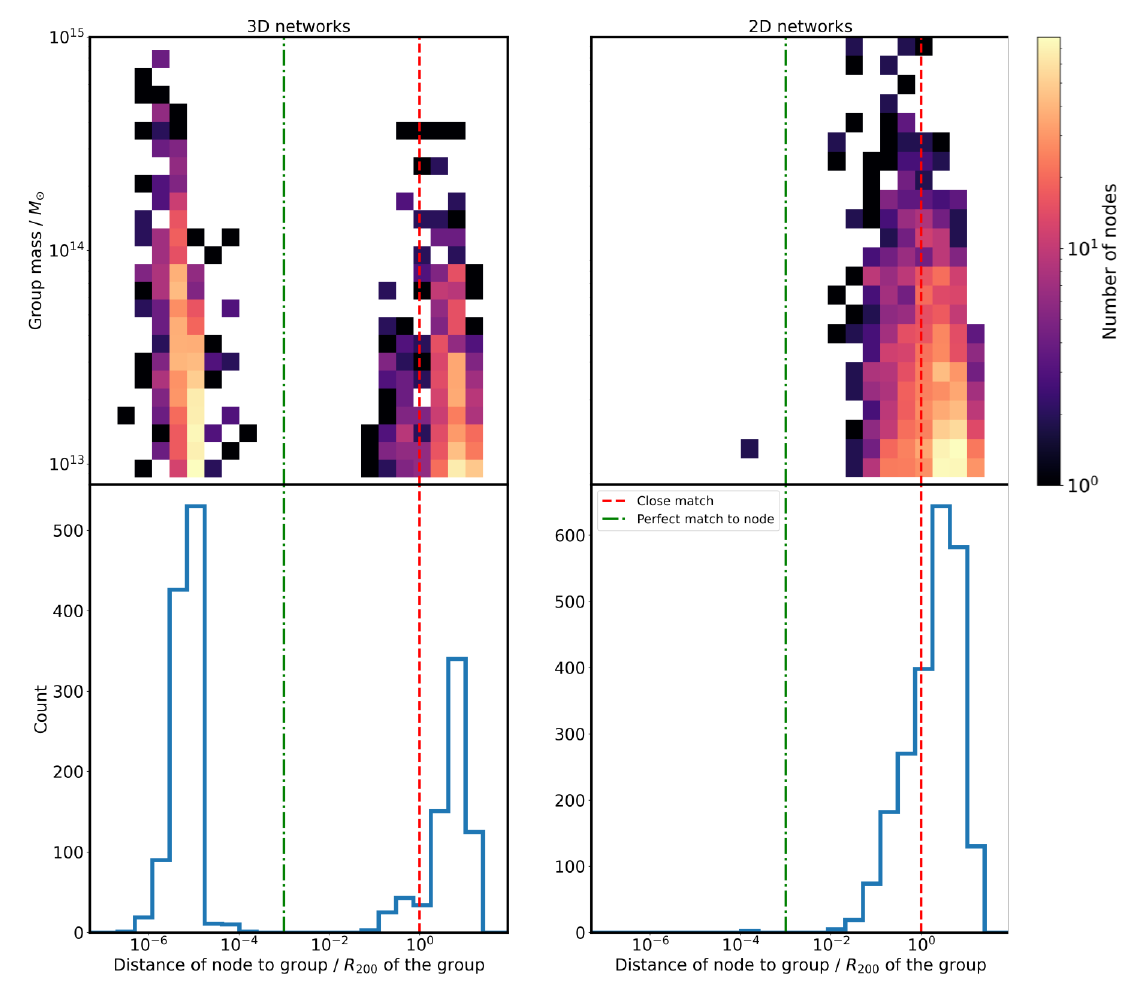}
    \caption{An illustration of the level of success between the matching of cosmic web nodes and galaxy groups in the 3D reference simulation (left) and the 2D projections (right). Top two panels: a colour map showing the normalized distance (in units of $R_{200}$ of each node to its nearest group against the mass of the group in the 3D simulations (left) and the 2D projections (right). The colour corresponds to the density of points. The dashed red line represents our criteria for a close match and the dot-dashed green line corresponds to a perfect match. The bottom two panels show the corresponding 1D histogram of the distance of each node to a group. In 3D, most cosmic web nodes successfully match to a galaxy group. However, the link is significantly weakened in 2D.}
    \label{fig:distance_node_to_group_MW}
\end{figure*}

\begin{figure*}
    \centering
    \includegraphics[width = \textwidth]{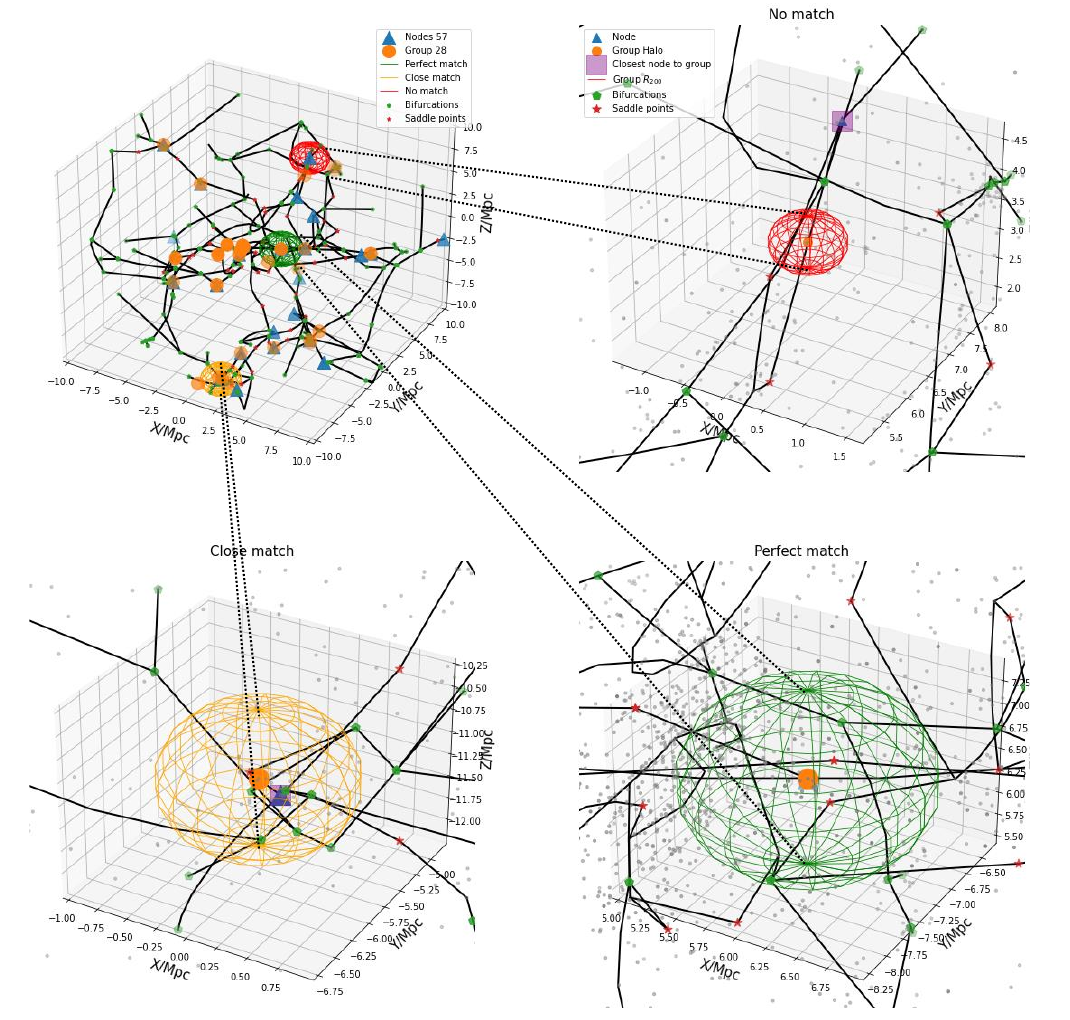}
    \caption{An example galaxy cluster ($M_{200} = 4.3\times 10^{14} M_{\odot}$) with the \texttt{DisPerSE} critical points and group haloes overlaid. The top left panel shows the full smoothed filament network in black. The cosmic web nodes are shown as blue triangles, the bifurcation critical points are displayed as green pentagons, and the saddle points are illustrated as red stars. We have also displayed the galaxy groups as orange circles for reference. As explained in the text and demonstrated in Figure~\ref{fig:distance_node_to_group_MW}, we calculate the distance between each node and each galaxy group within the cluster and show an example of a non-match, a close match and a perfect match as red, yellow, and green mesh spheres respectively. The top right panel illustrates a zoom in on a `non-match' between a node and a galaxy group, where we have also plotted the haloes from the simulation. The bottom left panel is a zoom in on an example of a `close match' and the bottom right panel shows an example of a `perfect match'. In the top right and lower left panels we have plotted the closest node to a group, showing where a cosmic web node has not latched on to the closest group. The radius of the mesh spheres in the upper right and lower two plots correspond to $R_{200}$ of the group halo.}
    \label{fig:example_cluster_matching}
\end{figure*}

\begin{figure}
    \centering
    \includegraphics[width = 0.33\textwidth]{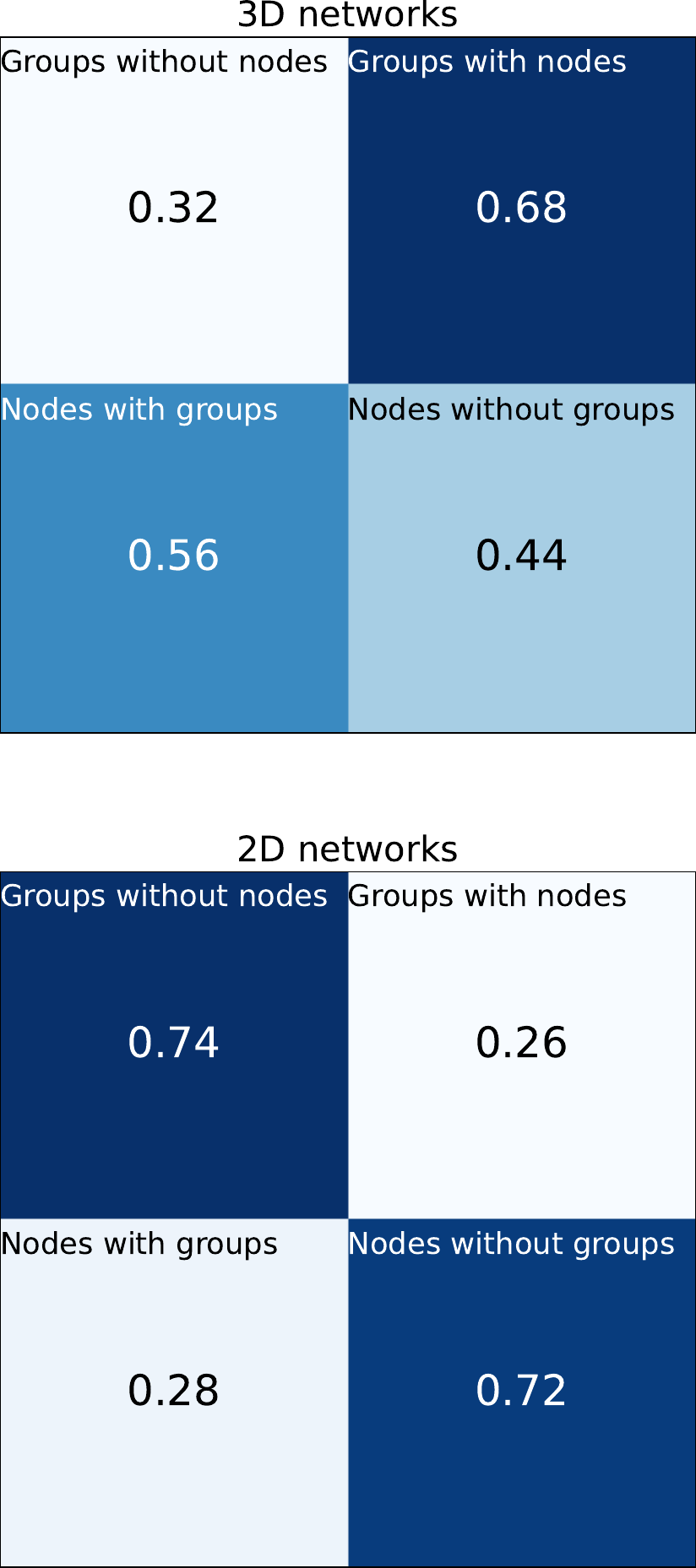}
    \caption{A confusion matrix illustrating the relative success of the matching of cosmic web nodes to galaxy groups in the 3D simulations (top panel) and the 2D projections (lower panel). The bottom row of each matrix is calculated by the number of nodes with/without groups divided by the total number of nodes. The top row is calculated as the number of groups with/without nodes divided by the total number of groups. We only consider groups and nodes in the region $1.5R_{200} < r < 5.5R_{200}$.}
    \label{fig:confusion}
\end{figure}

\begin{figure}
    \centering
    \includegraphics[width = 0.48\textwidth]{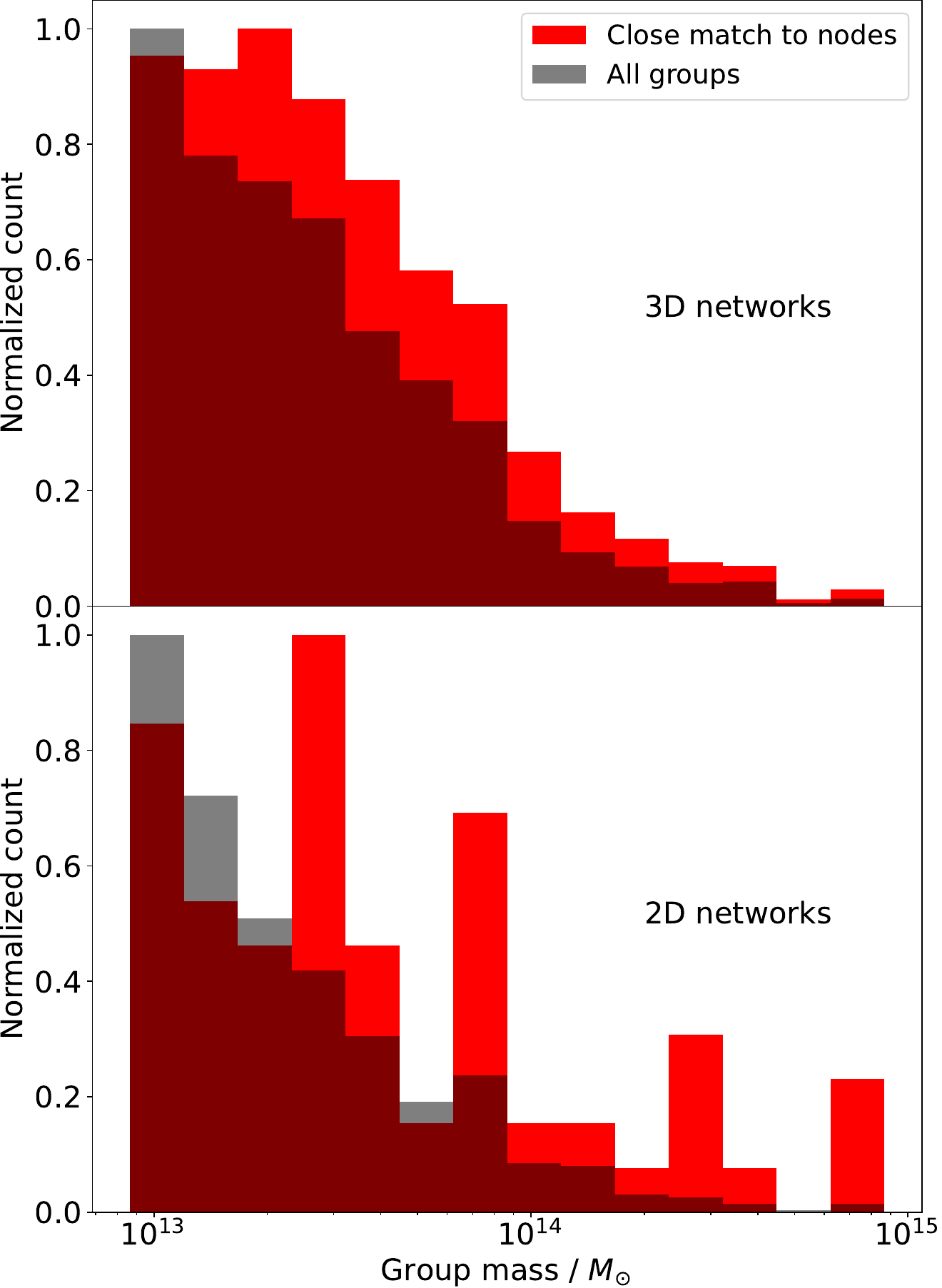}
    \caption{Mass distribution of all galaxy groups and those groups considered successful matches to nodes. We present the results for the 3D simulated clusters in the top panel and the results for the 2D projections in the lower panel. We display the normalized histograms of the group masses in black and the histogram for all galaxy groups that are close matches to nodes in red. The mass distributions appear significantly different, such that in the 3D reference simulations, groups that are close matches to nodes tend to be more massive than the general group sample.}
    \label{fig:mass_dist}
\end{figure}

Figure~\ref{fig:distance_node_to_group_MW} quantifies the success in matching cosmic web nodes to galaxy groups in the mass-weighted case. The top left panel of Figure~\ref{fig:distance_node_to_group_MW} shows the relationship between distance of a node to a galaxy group and the group mass, represented by a color map. The green dot-dashed vertical line represents the upper limit of what we label a `perfect match'. Anything considered a `perfect match' is where the distance between a node and a group is essentially zero, i.e.,  below the simulation resolution limit. The dashed red line illustrates the boundary of what we consider to be a close match to a group: this corresponds to $R_{200}$ of the respective group that we are analyzing.
There are two main peaks in the distribution, as replicated in the lower left panel, with the dominant peak in the `perfect match' range and the secondary peak corresponding to a slightly lower instance of non-matches. For the 3D simulated clusters, out of the 1818 nodes, 1011 are associated with galaxy groups ($56\%$). 
The lack of node--group pairs  in Figure~\ref{fig:distance_node_to_group_MW} with distances in the  $10^{-4} \lesssim R/R_{200} \lesssim 10^{-1}$ range is due to mass-weighting and the fact that nodes are always located at the centre of a halo: if a less massive halo resides very close to a group-mass halo ($R/R_{200} \lesssim 10^{-1}$) the node would `latch' to the group-mass halo itself, and not the lower-mass halo.

For illustration, in Figure~\ref{fig:example_cluster_matching} we display the filament network from one model galaxy cluster in our sample. In the top left panel we show the network in black, with the nodes as blue triangles and groups as orange discs. We also show the other \texttt{DisPerSE} critical points, saddle points (local density minima) and bifurcations (where two or more filaments intersect without a maximum being present). The other panels then zoom in on three different regions that exemplify a `no match' between a node and a group, a `close match' and an instance of a `perfect match', enclosed by a red, yellow and green mesh sphere. In our example in the top right panel, \texttt{DisPerSE} does not place a node where a group is. Instead, a node is identified at a distance of 1.75Mpc. In the bottom left panel, the node has been associated with a halo that is a subhalo of a galaxy group, but not the main group halo, and it is therefore a successful match but not a perfect match. This corresponds to the data points between the dash-dotted green line and the dashed red line in Figure~\ref{fig:distance_node_to_group_MW}. This demonstrates that while  mass-weighting helps, it does not always result in a direct match from the galaxy group halo to a node. In the bottom right panel, the cosmic web node has latched on to a group-sized halo. 

We quantify the success of matching nodes and groups using a confusion matrix in Figure~\ref{fig:confusion}. Summing over all of the clusters, we calculate a purity of $56\%$ and a completeness of $68\%$. Whilst we have statistically demonstrated that there is a link between the positions of cosmic web nodes and galaxy groups, we note that there remains significant contamination, with $44\%$ of nodes not matching to groups and $32\%$ of groups not matching to nodes. \footnote{We note that the purity and completeness are largely influenced by the persistence. By increasing the persistence, there will be less critical points and therefore, less nodes. In turn, this would decrease the completeness but increase the purity.}

In the top panel of Figure~\ref{fig:mass_dist} we show the normalized mass distribution of the entire sample of galaxy groups as well as the mass distribution of galaxy groups that are close matches to nodes. More massive groups are more likely to match \texttt{DisPerSE} nodes. To quantify this, we perform a Kolmogorov-Smirnov test, which is a non-parametric test of the equality of two continuous, one dimensional probability distributions. We test the null hypothesis that the cumulative mass distribution of groups could be drawn from the `close match' to node group mass distribution and chose a significance threshold of 0.05. We record a p-value of $10^{-6}$ which allows us to reject the null hypothesis. In other words, galaxy groups that are located at cosmic web nodes have an intrinsically different mass distribution to that of the general group population: they are typically more massive. This result agrees with \cite{Cohn22}. They further found that matched clusters to nodes tend to occur in nodes of higher density (correlated with cluster/group mass, see below), and have a slightly less recent major merger. This is something we will explore in future work.

\subsection{Matching groups to nodes in 2D projections}
\label{sec:Nodes_2D}
We expect that matching nodes to groups is more challenging in projected 2D. One of the obvious reasons is that we are losing 1/3 of the spatial information when we project the simulated cluster volumes. On the other hand, it is possible that in the 2D projections, we may produce false matches between nodes and groups. This is where a node is close to a group `on the sky', meaning they are a match in the 2D projections, but their real line-of-sight distance is large and would result in a non-match in the 3D simulations. With this in mind, in this section we investigate the success and limitations of the matching of cosmic web nodes to galaxy groups in 2D projections.

The right panels of Figure~\ref{fig:distance_node_to_group_MW} illustrate that the matching of cosmic web nodes to galaxy groups is significantly different compared to the 3D reference simulations. There is one main peak in the distance distribution that straddles the boundary of a `close match' but lies preferentially in the `no match' region. This is echoed in the bottom panel of Figure~\ref{fig:confusion}, where we present the purity and completeness. Overall, we find that of the 2327 cosmic web nodes, 662 of them match to galaxy groups (a purity of $28\%$). The corresponding completeness is significantly worse than the 3D case and is calculated to be $26\%$ compared to 68\%. In the bottom panel of Figure~\ref{fig:mass_dist}, we display the mass distributions of the galaxy group sample and the mass distributions of those groups that are close matches to nodes. We perform the same KS test, using the same null hypothesis and significance threshold and calculate a p-value of $10^{-5}$. We note that the mass distribution of groups here is contaminated by the projection into 2D and may add to the spurious peaks seen in the distribution.

As expected, the matching between nodes and groups in 2D returns a lower purity and completeness than in the 3D simulations. The dimensional reduction severely impacts the success in matching. However, Figure~\ref{fig:radial_groups_nodes} shows a convergence in the number density of nodes and groups at greater distances from the cluster core. With this in mind, in the next section we examine the matching of cosmic web nodes to galaxy groups as a function of clustercentric distance.

\subsection{Radial dependence on matching nodes to groups}

\begin{figure*}
    \centering
    \includegraphics[width = 0.9\textwidth]{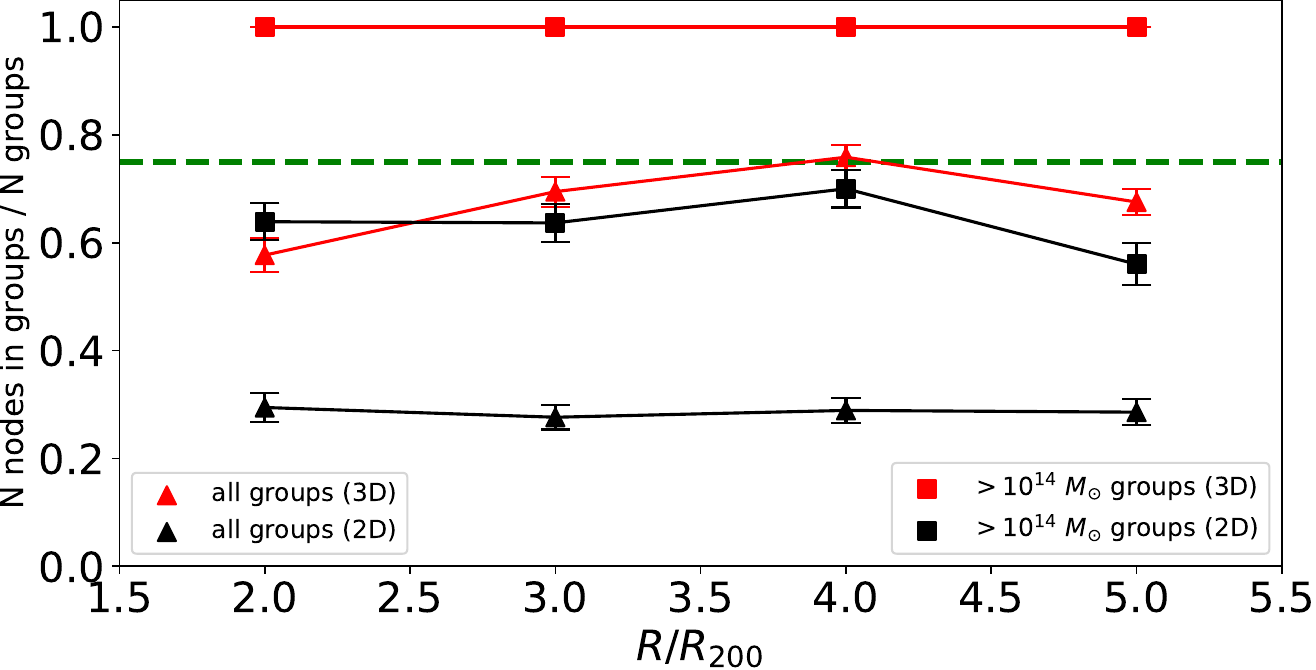}
    \caption{Fraction of groups in nodes (completeness) as a function of clustercentric radius, summed over our entire 3D simulated cluster sample and the 2D projected sample. We evaluate this fraction at four intervals from $1.5R_{200}$ to $5.5R_{200}$. The red triangles display the completeness in matching the entire group sample in 3D and the black triangles illustrate the completeness in 2D. We show the case where we limit to the most massive galaxy groups as red squares in 3D and in black squares in 2D. Error bars indicate the standard error. The green dashed line shows the average fraction obtained over much larger scales \citep{Cohn22} for comparison. In 3D, cosmic web nodes match well to very massive galaxy groups, with a success rate close to $100\%$.}
    \label{fig:radial_success_mw}
\end{figure*}

\subsubsection{3D filament networks}
\label{sec:3D_fil_radial}
We start by testing the radial matching of cosmic web nodes to galaxy groups using the same radial bins discussed in \ref{sec:3D_filaments} and present the findings in Figure~\ref{fig:radial_success_mw}. Here, the y axis is a measure of the fraction of cosmic web nodes that reside in galaxy groups divided by the total number of groups in that radial bin -- the completeness. This is calculated for each cluster and the mean is represented by the red triangles with the error bars representing the standard error. 

Generally, for the entire group catalogue, the success rate improves as we increase the clustercentric distance. This is to be expected -- close to the cluster core the main halo dominates the density field and therefore prevents the formation of critical points that exceed our persistence threshold. As the gravitational influence of the cluster decreases, more persistence pairs can form and are therefore more likely to align with the high mass groups. Encouragingly, as the clustercentric distance increases, the success rate approaches that of \cite{Cohn22}, although we must bear in mind that the mass range of the haloes we use ($10^{11} $ M$_{\odot}$ $< M_{\text{halo}} < 10^{15}$  M$_{\odot}$) is much larger than the one used by this author ($M > 10^{14}$  M$_{\odot}$); furthermore, we are probing very different distance scales (tens of Mpcs compared to hundreds of Mpcs). Nevertheless, our relatively high success at matching nodes and groups in the complex vicinity of clusters (at least in 3D) seems promising. We note that the  mass-weighting scheme we use in \texttt{DisPerSE} plays a very important role (cf. Appendix~A) and is largely responsible for the success in matching nodes to galaxy groups.

In Figure~\ref{fig:radial_success_mw} we also display the success rate for massive galaxy groups, using the same mass threshold as the \cite{Cohn22} work ($M>10^{14}$ M$_{\odot}$). Limiting the sample to these high mass groups, we find that completeness jumps to $100\%$. This implies that the derivation of \texttt{DisPerSE} nodes can be used in the detection of nearly all massive galaxy groups in close proximity to clusters when one applies mass-weighting.

\subsubsection{2D filament networks}
We repeat the analysis for the 2D projected networks. Contrary to the results of Section~\ref{sec:3D_fil_radial}, we find that there is no improvement in the success rate with increasing radius, resulting in a flat rate of approximately $26\%$. However, when we consider only the higher mass groups, we find a stark improvement in the matching of nodes to galaxy groups: $63\%$ of $M > 10^{14}$  M$_{\odot}$ groups match to a cosmic web node. Interestingly, there appears to be little dependence on the clustercentric distance.

We conclude that the effect of the large contamination rates in the purity and completeness of the node-matched group sample prevent us from using \texttt{Disperse} to identify a robust sample of galaxy groups. However, we have shown that our approach is much more successful when considering only the most massive groups. Therefore, we expect to be able to identify $\sim63\%$ of all galaxy groups with $M > 10^{14} M_{\odot}$ using nodes identified by \texttt{DisPerSE}. Furthermore, for all group masses, \texttt{DisPerSE} can be applied to observations to locate potential galaxy groups using cosmic web nodes, and then the resulting sample can be verified and cleaned using alternative, perhaps more ad-hoc and less systematic methods. 

\section{Group mass estimation from cosmic web node density}
\label{sec:group_mass_node_dens}
\begin{figure}
    \centering
    \includegraphics[width = 0.45\textwidth]{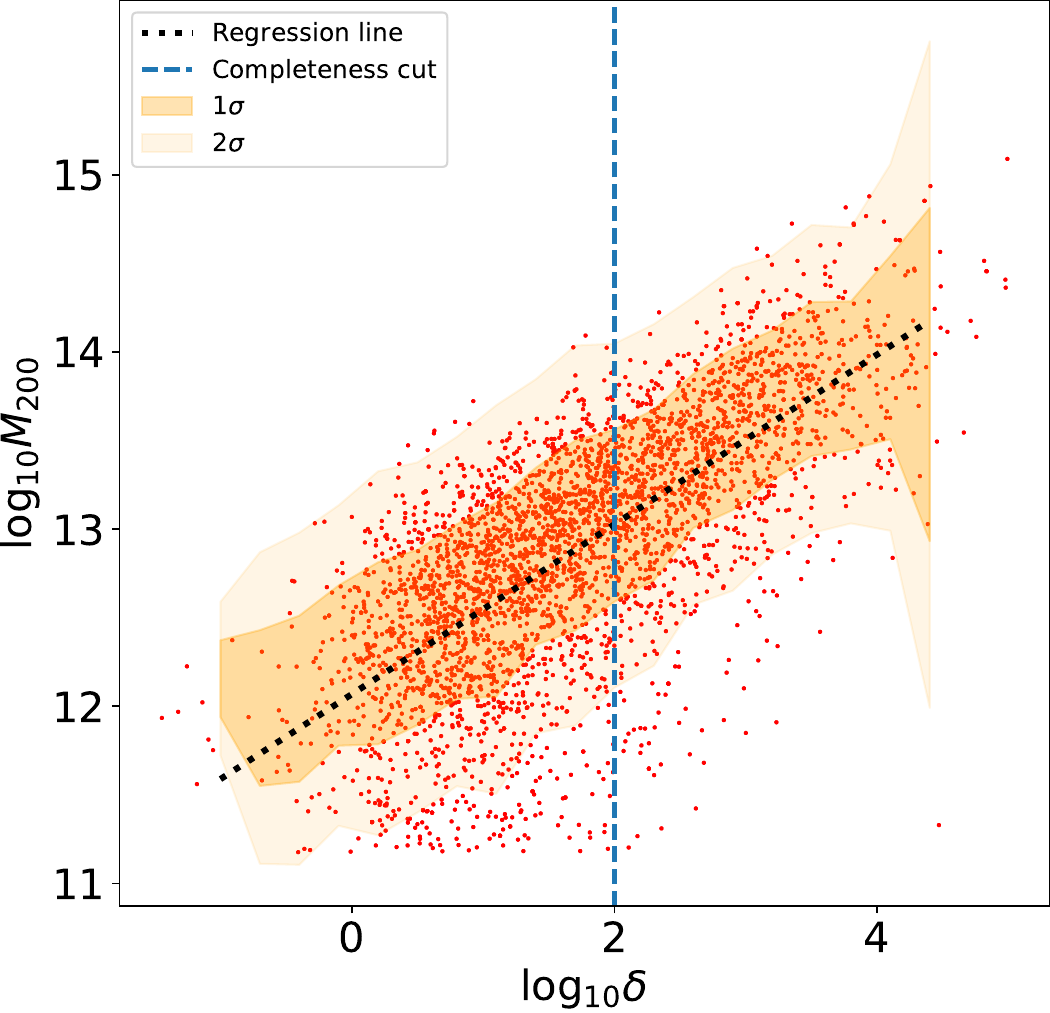}
    \caption{Correlation between the mass of a galaxy group and the density of the node that it matches to in the 3D reference clusters (red dots; see text for details). The dotted black line shows the regression line obtained by fitting only the data in the complete sample, as discussed in the text. The sample is complete for $\log_{10}{\delta} > 2$  (indicated by the blue dashed line). The $1\sigma$ and $2\sigma$ scatter are overlaid using two different shades of yellow. The node density of the cosmic web can be used to estimate the mass of a galaxy group, albeit with large scatter.}
    \label{fig:groupmass_approx}
\end{figure}
We have shown that it is possible to locate a significant fraction of galaxy groups using cosmic web nodes in 3D. We now examine whether we can use \texttt{DisPerSE} to estimate the masses of these groups. In what follows, we demonstrate that the node density of the cosmic web provides information on the masses of galaxy groups that are associated with them. 

As mentioned in Section~\ref{sec:3D_filaments}, \texttt{DisPerSE} outputs a list of density values that are calculated during the Delaunay tessellation for each identified critical point. We have demonstrated that, in 3D, there is a tendency for cosmic web nodes to match to galaxy groups and we therefore further investigate the possibility of using the density of the nearest node to a galaxy group in order to estimate its mass. We do this by taking the galaxy groups that are a `perfect match' to a node and record its corresponding node density. In Figure~\ref{fig:groupmass_approx} we show that there is a strong positive correlation between these two parameters, albeit with considerable scatter. 

We then fit a least-squares regression line to the sample of nodes for which a complete mass-selected sample of corresponding groups can be identified in the simulations.  Visual inspection of Figure~\ref{fig:groupmass_approx} indicates that groups with masses that correspond to node densities of $\log_{10} \delta > 2$, indicated by the blue dashed line in the figure, constitute such a complete sample. We only fit a regression line to that complete sample to avoid any Malmquist-like biases. The figure shows that an extrapolation of this line to lower group masses is a fair representation of the trend at all masses. The equation of the regression line is 
\begin{equation}
  \log_{M_{200}} = 0.48\log_{10}\delta + 12.07.
\end{equation}
Using this regression line it is possible to obtain a rough estimate for the mass of the groups selected as \texttt{DisPerSE}-identified nodes, but the scatter is large, roughly a factor of $\sim3$.

\subsection{High mass groups and node density matching}
We have shown that there is a strong positive correlation between the node density and mass of its closest matched halo. It is therefore possible to select the highest mass groups ($M > 10^{14}$ M$_{\odot}$) by finding a suitable cut in the node density. In this section, we test the matching of high mass groups to high density nodes in order to uncover whether this method is a robust strategy for detecting massive galaxy groups.

Using equation~3 we find that a group mass of $10^{14}$ M$_{\odot}$ corresponds to $\delta \sim 10000$ in the node density. By considering only the highest mass groups and the nodes above this density threshold, we repeat the analysis in Section~\ref{sec:Nodes_3D} by quantifying the matching of high mass groups to high density nodes. We note that we exclude clusters from this analysis that do not match to any high mass galaxy groups. 
\begin{figure}
    \centering
    \includegraphics[width = 0.4\textwidth]{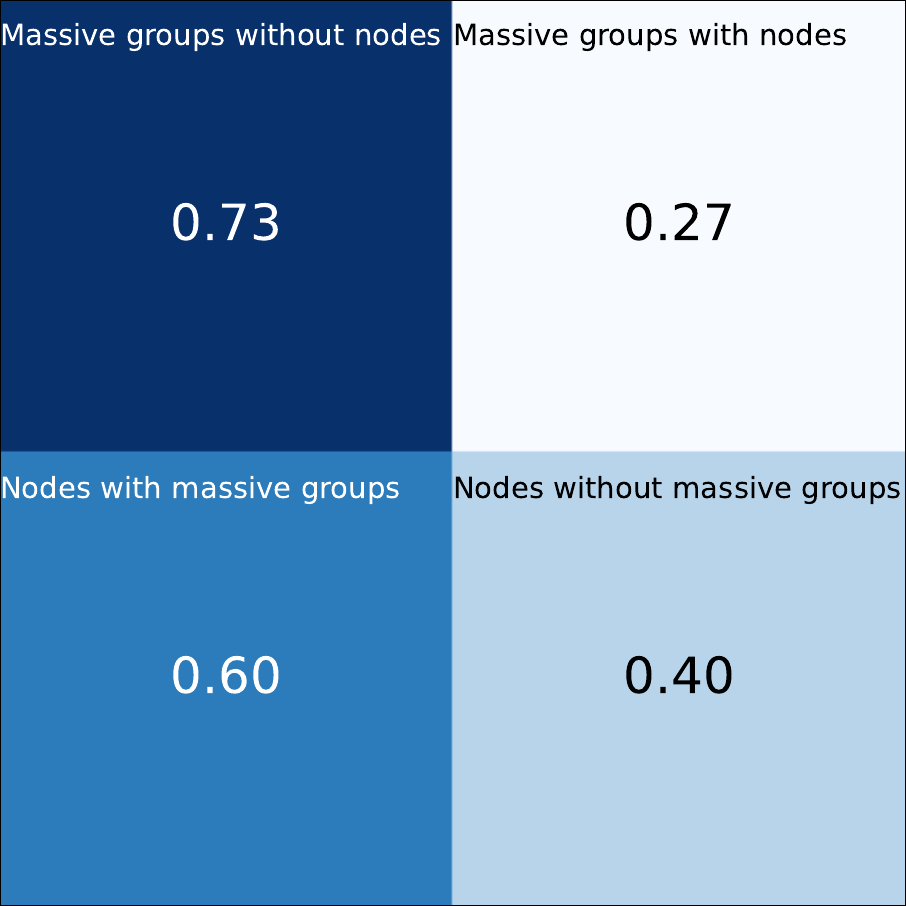}
    \caption{Same as Figure~\ref{fig:confusion} for the 3D reference simulations but for high mass groups ($M > 10^{14} M_{\odot}$) selected as high density nodes, as described in Section~\ref{sec:group_mass_node_dens}.}
    \label{fig:confusion_highmass}
\end{figure}

Figure~\ref{fig:confusion_highmass} illustrates the results from this analysis. We find that the fraction of high-density nodes in high mass galaxy groups (purity) is 60\%, very similar to that of the entire group sample. However, the fraction of high mass groups matching high density nodes (completeness) decreases to 27\%. We attribute this to there being a greater number of groups (89) than nodes (42) above the mass and density thresholds.  When we limit the node density, we exclude some cases where nodes match with galaxy groups but lie below the density threshold, thereby negatively affecting the matching of massive groups to high-density nodes. We conclude that, although there is a strong correlation between the group mass and cosmic web node density, restricting the node density in this way does not significantly improve purity or completeness.

\section{Conclusions}
\label{sec:Conclusions}
Galaxies experience different physical processes in different cosmic web environments. Next generation wide-field spectroscopic surveys will, for the first time, be able to accurately map the distribution of galaxies to cosmic web features around statistical samples of galaxy clusters, where pre-processing is present. In this paper, we present and evaluate a novel approach for identifying galaxy groups (haloes with $\sigma_{v} > 300 h^{-1}$ km/s) near massive galaxy clusters utilizing the critical points identified as network nodes using the \texttt{DisPerSE} software, (which we run on haloes with more than 100 dark matter particles). We summarize our main findings below.

\begin{enumerate}
    \item We have tested the matching of cosmic web nodes, (derived from mass-weighted filament networks from \texttt{DisPerSE}), to galaxy groups in and around massive galaxy clusters in 3D and projected 2D cluster simulations. We find a purity of $\sim 56\%$ and a completeness of $\sim 68\%$ in 3D and a purity of $\sim 28\%$ and a completeness of $26\%$ in 2D. 
    \item We find that the galaxy groups that closely match with cosmic web nodes tend to be the more massive ones. 
    \item In the 3D reference simulations, we find a slight improvement in the fractional number of nodes within galaxy groups as we move further away from the cluster core. This suggests that the cluster core's complexity hinders the accurate matching of density field maxima to galaxy groups. Conversely, in the cluster outskirts the success rate increases due to the dominant influence of the large scale cosmic web rather than the cluster core. Within the range of 3--5$R_{200}$, the number of nodes in groups reaches a maximum of approximately $\sim 75\%$, matching the results obtained by \cite{Cohn22} over much larger spatial scales. In contrast, in the 2D projections, we do not see any radial trend.
    \item Limiting our analysis to only the most massive galaxy groups $(M>10^{14} M_{\odot})$, we find that $100\%$ of cosmic web nodes match to galaxy groups in the reference 3D simulations. We also find a stark improvement in the success rate for 2D projections, increasing from $\sim28\%$ to $\sim63\%$.
    \item We find a strong positive correlation between the mass of the groups and the \texttt{DisPerSE}-determined density of their matching nodes. This correlation (equation~3) can be used to obtain a rough estimate of the group mass within a factor of $\sim3$. 
\end{enumerate}
In summary, we have shown that the widely-used topological filament finder \texttt{DisPerSE} can be used as a powerful tool for identifying galaxy groups around clusters. It can be further complimented with other group finding algorithms. We have tested the strengths and weaknesses of this approach with future wide-field surveys of galaxy clusters in mind. While we used simulated galaxy clusters in this study, its accuracy can be scrutinized with diverse group-finding methods and observational data, thus opening new avenues for the study of galaxy groups and their role in galaxy evolution.

\section*{Acknowledgements}
DJC, AAS, UK, and MEG acknowledge financial support from the
UK Science and Technology Facilities Council (STFC; through a PhD studentship, and consolidated grant ref
ST/T000171/1). WC is supported by the STFC AGP Grant ST/V000594/1 and the Atracci\'{o}n de Talento Contract no. 2020-T1/TIC-19882 granted by the Comunidad de Madrid in Spain. He also thanks the Ministerio de Ciencia e Innovación (Spain) for financial support under Project grant PID2021-122603NB-C21 and ERC: HORIZON-TMA-MSCA-SE for supporting the LACEGAL-III project with grant number 101086388. AK is supported by the Ministerio de Ciencia e Innovaci\'{o}n (MICINN) under research grant PID2021-122603NB-C21 and further thanks Led Zeppelin for immigrant song.

For the purpose of open access, the authors have applied a creative commons attribution (CC BY) to any journal-accepted manuscript.

This work has been made possible by \texttt{TheThreeHundred} collaboration, which benefits from financial support of the European Union’s Horizon 2020 Research and Innovation programme under the Marie Sk\l{}odowskaw-Curie grant agreement number 734374, i.e. the LACEGAL project. \texttt{TheThreeHundred} simulations used in this paper have been performed in the MareNostrum Supercomputer at the Barcelona Supercomputing Center, thanks to CPU time granted by the Red Espa\~{n}ola de Supercomputaci\'{o}n. 

%%%%%%%%%%%%%%%%%%%%%%%%%%%%%%%%%%%%%%%%%%%%%%%%%%
\section*{Data Availability}
Data available on request to \texttt{TheThreeHundred} collaboration: https://www.the300-project.org.


\bibliographystyle{mnras}
\bibliography{bibliography} % if your bibtex file is called example.bib

\appendix
\section{Non mass-weighted networks}
\begin{figure*}
    \centering
    \includegraphics[width = \textwidth]{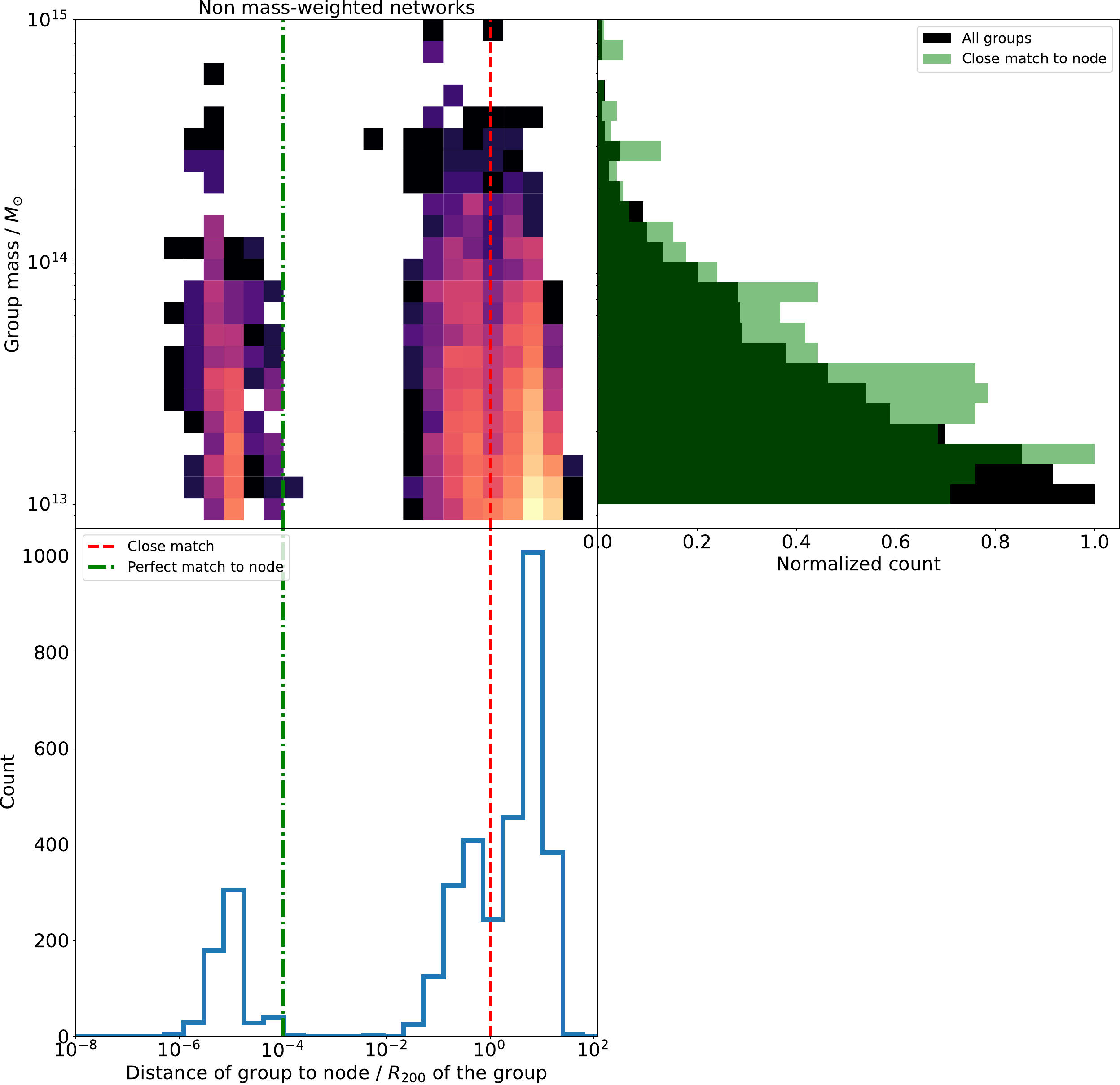}
    \caption{The same plot as Figure~\ref{fig:distance_node_to_group_MW} but for the non mass-weighted case, completed in 3D. In the right panel we show the mass distributions of the entire group sample and compare it to the sample of groups that are close matches to nodes, as done in Figure~\ref{fig:mass_dist}.}
    \label{fig:distance_node_to_group_NMW}
\end{figure*}

For comparison, we repeat the process of determining the distance between each cosmic web node to its nearest galaxy group but without mass-weighting. Overall, the matching is much less successful than in the mass-weighted case: without mass-weighting we only find $43\%$ of the cosmic web nodes match galaxy groups. We demonstrate this in Figure~\ref{fig:distance_node_to_group_NMW}. The presence of a third peak, seen in the lower left histogram at distances $R/R_{200}\sim10^{-1}$, is due to the node latching on to a subhalo within the group halo. This result implies using mass-weighting in the filament finding process very significantly improves our ability to locate galaxy groups using network nodes. 

%\begin{thebibliography}{99}
%\end{thebibliography}

%%%%%%%%%%%%%%%%%%%%%%%%%%%%%%%%%%%%%%%%%%%%%%%%%%

%%%%%%%%%%%%%%%%% APPENDICES %%%%%%%%%%%%%%%%%%%%%

%%%%%%%%%%%%%%%%%%%%%%%%%%%%%%%%%%%%%%%%%%%%%%%%%%

% Don't change these lines
\bsp	% typesetting comment
\label{lastpage}
\end{document}